# The symmetry-adapted configurational ensemble approach to the computer simulation of site-disordered solids

**Ricardo Grau-Crespo[1],* and Said Hamad [2]**

[1]  Department of Chemistry, University of Reading, Whiteknights, Reading RG6 6AD, United Kingdom.
[2]  Departamento de Sistemas Físicos, Químicos y Naturales, Universidad Pablo de Olavide, Carretera de Utrera km. 1, 41013 Seville, Spain.
*  Author to whom correspondence should be addressed; E-Mail: r.grau-crespo@reading.ac.uk; Tel.: +44 118-378-7180

**Abstract:** Site-occupancy disorder, defined as the non-periodic occupation of lattice sites in a crystal structure, is a ubiquitous phenomenon in solid-state physics and chemistry. Examples are mineral solid solutions, synthetic non-stoichiometric compounds and metal alloys. The experimental investigation of these materials using diffraction techniques only provides averaged information of their structure. However, many properties of interest in these solids are determined by the local geometry and degree of disorder, which escape an "average crystal" description, either from experiments or from theory. In this paper, we describe a methodology for the computer simulation of site-disordered solids, based on the consideration of configurational ensembles and statistical mechanics, where the number of occupancy configurations is reduced by taking advantage of the crystal symmetry of the lattice. Thermodynamics and non-thermodynamic properties are then defined from the statistics in the symmetry-adapted configurational ensemble. We will briefly summarize and discuss some recent applications of this methodology to problems in materials science.

## 1. Introduction

There are many problems in materials science involving the solution of two or more crystalline materials with some level of randomness in the occupancy of lattice sites. We define site disorder as the kind of disorder that results from *non-periodic* occupation of lattice sites in a crystal structure. Amorphous disorder differs from lattice site disorder in that the latter does not destroy the long-range periodicity of the lattice sites, except possibly with small local atomic displacements with respect to lattice sites. Examples of site-disordered materials include metallic alloys, mineral solid solutions, and synthetic non-stoichiometric compounds. The computational modeling of these solids is challenging because periodic boundary conditions cannot be applied in the same straightforward way as in the simulation of ordered crystals.

## 2. Classification of site-disorder models

The models employed in the literature to investigate site-disordered solids can be classified in three broad groups (**figure 1**).

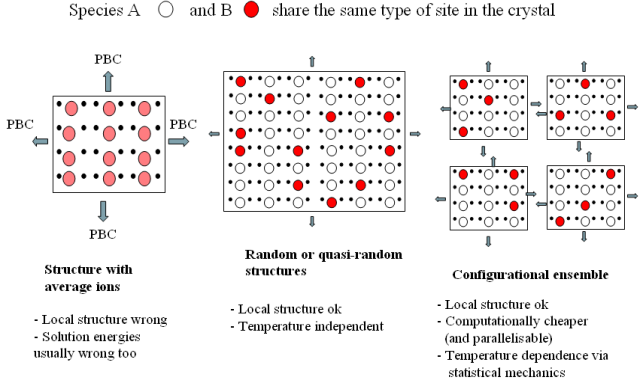

**Figure 1.** Classification of methods to represent site-disorder in solids

The first group comprises all methods in which a sort of average atom is defined, thus allowing recovering the perfect periodicity of the crystal. In the context of classical calculations, based on analytical interatomic potentials, this is done by making each site experience a potential which is the mean or weighted average of all possible configurations corresponding to disordered atomic positions. This approach is implemented, for example, in the GULP code [1,2], and can sometimes be useful for preliminary simulations of very disordered (random) systems. An equivalent method in the world of quantum-mechanical simulations is the virtual crystal approximation (VCA), where the potential felt by electrons is the one generated by average atoms, *i.e.* average of potentials of atoms that can occupy a given site and its periodic images. The main drawback of this kind of methods is that the local structure around each particular ion in a real material is very poorly represented by the geometry around these average ions.

In the second group of methods for treating site-disordered solids, a large periodic supercell is employed with a more or less random distribution of ions at the sites. This type of representation is computationally more expensive than the "average-ion" models, but it provides a better description of the local geometries found in the real system. It is assumed that (a) distribution of atoms on lattice sites is random, and (b) the supercell is large enough to include a large number of possible local arrangements of ions, amounting to spatial averaging over configurations. A useful variation of this model is the special quasi-random structure, where the ion positions in the supercell are chosen to mimic as closely as possible the most relevant near-neighbor pair and multisite correlations of a random substitutional alloy [3]. Special quasi-random structures are particularly useful in evaluation of the electronic structure and related properties such as magnetic moments of site-disordered solids. Their main limitation comes from the inflexibility of the ion distribution in the structure, which is fixed to mimic a disordered (a truly random) solid. However, we often desire to investigate varying degrees of disorder, for example, short-range ordering can be present depending on the temperature used in the synthesis, for which we need a more flexible representation.

The third type of methods is the multi-configurational supercell approach, which is the focus of the present paper. Within this approach, an infinite site-disordered solid is modeled with a set of configurations with of various site occupancies in a supercell representing a piece of the solid. Each configuration corresponds to a particular arrangement of the atoms within the supercell, and has associated a probability of occurrence. The idea behind the method is that listing all possible configurations and their probabilities can provide a reasonable description of the distribution of the ions and their level of disorder, at least within the range marked by the supercell size.

In order to attribute experimental meaning to this kind of representation, we can image the case of a site-disordered surface that is being studied with an electron microscope, capable of taking atomic resolution photographs of the surface. If we take a very large number of photographs at randomly selected sections of the surface, we have all the information required to describe any distribution pattern with ordering range shorter than the size of the photographs. The probability of a given configuration can then be defined as its frequency of appearance in the limit of a very large number of images. In the limit of infinite size of the system, this scheme becomes exact.

From a theoretical point of view, the central challenge in the multi-configurational supercell approach is the calculation of the probabilities of occurrence of the configurations, under the assumption of configurational equilibrium. A typical approximation consists of assigning an energy to each configuration, and then applying a formalism based on Boltzmann-Gibbs statistical mechanics, in such a way that those configurations with lower energies have higher probabilities, and the dispersion of the distribution is controlled by the temperature: at low temperatures only the most stable configurations occur, whereas at high temperatures more configurations are accessible to the ions, leading to higher degree of disorder.

We should note that the existence of a well-defined energy characterizing the stability of each configuration is not trivial, as strictly speaking the energy contribution from a given ionic configuration in a cell depends on the configuration of the ions in the neighboring cells. Only for large supercells, where *intracell* interactions are much more important than *intercell* interactions, the energy of a given configuration can be considered a function of the arrangement of the ions within the supercell. In what follows we will consider that this is the case, i.e., the energy of a given configuration is independent of the distribution of cations next to the cell boundary. Then, we could as well assume that the distribution of ions in the region adjacent to supercell is identical to that in the supercell used in simulations. Thus, we can simply calculate the configuration energy using periodic boundary conditions, which is straightforward using modern computer programs for solid state simulations.

Orthogonally to the classification of the methods in terms of the representation of disorder, we can also classify the methods in terms of the types of methods used for evaluating the energy of the periodic configurations in the crystal. This is actually a typical task in computational physics and chemistry, and the different methods and approximations involved have been widely discussed elsewhere (e.g. [4]). We only present here a simple classification in increasing order of sophistication and computational cost:

*Type I methods*: in this case the energy is a function of only the site occupancies. Electronic and structural (geometric) degrees of freedom are not included explicitly, but are implicit within a model that yields the energy from nearest-neighbor (NN), next-nearest-neighbor (NNN) configurations or longer-distance effective pair interactions, or including terms for clusters of more than two ions. These types of methods include Ising-type models of alloys and cluster expansion methods, and have been used extensively in determination of phase diagrams of alloys.

*Type II methods*: including geometric relaxations explicitly, but electronic effects only implicitly. The energy is evaluated in this case via a classical interatomic potential function or force field, which is a function of the ionic coordinates. These are quite efficient computationally and can be used in studies of symmetry-breaking structural phase transitions such as those in a ferroelectric or a shape memory alloy. In such a case, the structural degrees of freedom are directly relevant to the problem and cannot be integrated out.

*Type III methods*: they include explicit geometric and electronic relaxation for each configuration. This category comprises all quantum-mechanical methods, including those based on the density functional theory (DFT) and its extensions, Hartree-Fock (HF) and post-HF approaches, hybrid DFT/HF, semi-empirical methods like tight-binding, etc. Computationally, these methods are quite expensive and can be used only with relatively small supercells. They have to be used in problems that involve electronic phase transitions (such as magnetic or metal-insulator transition), and those where site-specific chemistry is relevant, for example in catalysts.

In the evaluation of energies of a large number of configurations to investigate the thermodynamics of disorder, type I methods are still the most commonly employed. Not only they are computationally cheaper, but they are also easier to integrate with sampling algorithms (e.g. Metropolis – Monte Carlo) within a single computer program. In contrast, energy evaluations using type II and type III methods are more expensive and typically require a specialized program that deals with only one geometric configuration at a time. Configurational sampling in these cases requires multiple calls to the quantum-mechanical or interatomic potential code from an external program, and has very long running times. However, there are good reasons to move towards more sophisticated (type II and III) methods to evaluate energies in a multi-configurational simulation. First, such calculations not only provide energies, but many other properties too (e.g. local geometries and cell parameters, and in the case of type III methods, electronic structure information). Any property that can be obtained for each configuration can be averaged over the ensemble to obtain effective values for the disordered solid. Second, methods of type II and III also provide access to vibrational properties of the solid and its response to external pressure, therefore allowing an integration of configurational and vibrational degrees of freedom in the construction of complex phase diagrams (as a function of composition, temperature and pressure). Finally, if interactions in the system are long-range, energy evaluations in terms of simple type I methods might not provide good precision, or would require a large number of terms and parameters [5]. The main disadvantage of methods of type II and III, while evaluating a large number of configurations, is their computational cost, but with the developments in computer hardware and efficient algorithms, they are becoming much more affordable.

## 3. Statistical formulation of the method

### 3.1 The Boltzmann equations

We now present the statistical formulation of the configurational equilibrium in a site-disordered binary system. For the sake of simplicity, in this initial formulation all configurations are constrained to have the same compositions, and we will ignore vibrational and pressure effects in the thermodynamics; the corresponding generalizations are introduced later in this chapter.

The extent of occurrence of the each configuration (labeled with an index $k$) is described in this approximation by a Boltzmann-like probability which is calculated from the energy $E_k$ of the configuration and the temperature $T$:

$$P_k = \frac{1}{Z} \exp(-E_k / k_B T)$$

(1)

where $k_B$ is Boltzmann's constant (it is formally equivalent to use the gas constant $R$ instead, and expressing the molar energies of supercells, but we follow here the usual notation in statistical mechanics in terms of $k_B$),

$$Z = \sum_{k=1}^{K} \exp(-E_k / k_B T)$$

(2)

is the partition function, and $K$ is the total number of configurations with the given composition in the supercell. For a binary system, $K$ can be calculated as a number of combinations:

$$K = \frac{N!}{(N-n)! n!}$$

(3)

where $N$ is the number of exchangeable sites in the supercell, and $n$ is number of ions of one of the species (it can also be a vacancy) that can occupy these sites. The molar concentrations are then $x=n/N$ for the first species, and $1-x$ for the second species.

The definition of the configurational ensemble and the associated probabilities allows us to obtain the effective value in the disordered solid of any quantity that can be theoretically obtained for each ordered configuration. If $A_k$ is the value of the given magnitude for configuration $k$, then the effective value in the disordered solid is:

$$A = \sum_{k=1}^{K} P_k A_k \qquad (4)$$

In this way, it is possible to obtain effective values even for quantities like the cell parameters, which are not strictly defined but still have experimental meaning in a solid with non-periodic distribution of ions on lattice sites. Equation (4) also allows us to obtain the effective energy of the solid from the configurational energies $E_k$, as:

$$E = \sum_{k=1}^{K} P_k E_k \qquad (5)$$

In evaluating the thermodynamic stability of a disordered solid at a given temperature, not only the energy but also the configurational multiplicity of the system should be taken into account, which is done by introducing the (configurational) free energy:

$$F = -k_B T \ln Z = -k_B T \sum_{k=1}^{K} \exp(-E_k / k_B T) \qquad (6)$$

The difference per temperature unit between the average energy and the free energy defines the configurational entropy, which can also be expressed in terms of the probabilities $P_k$:

$$S = \frac{E-F}{T} = -k_B \sum_{k=1}^{K} P_k \ln P_k \qquad (7)$$

We can distinguish two important limiting cases here:

*Perfect order:* this occurs when one configuration (say $k$=1) is much more stable than the rest, *i.e.*, it is separated from the other configurations by an energy difference much larger than $k_B T$. In this case $P_1$=1, while $P_k$=0 for $k \neq 1$, and the system will have zero configurational entropy. The configurational free energy simply corresponds to the energy of the most stable configuration.

*Perfect disorder*: this occurs when the energies of all the configurations are very similar (again in comparison with $k_B T$) or formally in the limit $T \rightarrow \infty$. In this case, all configurations have

the same probability $P_k$=1/$K$ and the configurational entropy reaches its maximum possible value:

$$S_{\max} = k_B \ln K = k_B \ln \frac{N!}{[N(1-x)]![Nx]!} \qquad (8)$$

which, in the limit of an infinitely large supercell ($N \rightarrow \infty$ at constant $x$), using Stirling's formula, converts to the well-known expression:

$$S_{\text{ideal}} = -k_B N (x \ln x + (1-x) \ln(1-x)) \qquad (9)$$

Intermediate to these two limiting cases, there is a continuum of situations with varying degrees of ordering, which can be described within the same formalism, leading to temperature-dependent entropy values given by Eq. (7). It is clear from the equations above that any finite supercell is unable to describe exactly the perfect disorder limit. In order to correct for this, it is convenient to re-write the free energy (Eq. 6) as:

$$F = -k_B T \ln K - k_B T \ln \left( \frac{1}{K} \sum_{k=1}^{K} \exp\left(-E_k / k_B T\right) \right) \qquad (10)$$

as suggested by Becker et al.[6]. The first term represents the entropy contribution in the limit of perfect disorder, while the second term contains the energy contribution plus the correction to the entropy contribution due to partial ordering. The first term can then be adjusted to its correct value, *i.e.*:

$$F = N k_B T \left( x \ln x + (1-x) \ln(1-x) \right)$$
$$-k_B T \ln \left( \frac{1}{K} \sum_{k=1}^{K} \exp(-E_k / k_B T) \right) \qquad (11)$$

which is equivalent to amending the temperature-dependent entropy by a term $\Delta S_{\text{corr}}$=$S_{\text{ideal}}$-$S_{\max}$, thus guaranteeing the correct behavior in the limit of perfect disorder. However, this adjustment also breaks down in the description of the perfect order limit, by introducing a spurious configurational entropy contribution to the free energy (which should be zero in this limit). Therefore this correction should be applied only to simulations of systems with nearly-perfect disorder.

## 3.2 Including vibrational contributions

This basic methodology can be made more sophisticated to include other effects, depending on the problem in hand. For example, in order to consider vibrational contributions to the thermodynamics of a disordered solid within the multi-configurational formalism, we can write the total partition function of the system as:

$$Z = \sum_{k=1}^{K} \sum_{v} \exp(-E_{k,v} / k_B T) \qquad (12)$$

where $v$ is an index (or strictly speaking, a collection of indices) that characterizes each vibrational state (with energy $E_{k,v}$) of configuration $k$. In terms of the vibrational partition function $Z_k^{(vib)}$ and the vibrational free energy $F_k^{(vib)}$ of each configuration, this becomes:

$$Z = \sum_{k=1}^{K} Z_k^{(vib)} = \sum_{k=1}^{K} \exp(-F_k^{(vib)} / k_B T) \qquad (13)$$

which, by comparison with Eq. (6), indicates that the formalism including vibrational contributions is equivalent to the one introduced in the previous section, except that now the vibrational free energy $F_k^{(vib)}$ of each configuration should be used to define the probabilities (1) instead of energy $E_k$ of the configuration. Using methods of type II and III, where the energy depends explicitly on ionic coordinates, vibrational free energies can be evaluated from the vibrational frequencies of each configuration, by invoking the harmonic approximation[7]. This, of course, adds considerably to the cost of the simulations, especially if the equilibrium geometry of each configuration is obtained by minimizing the free energy and not just the energy, but can become affordable if methods of Type II are being used (e.g. ref. [8]).

Analogously, if we want to introduce the effect of a finite external pressure $p$, we should use the Gibbs free energy:

$$G_k^{(vib)} = F_k^{(vib)} + pV_k = H_k - TS_k^{(vib)} \qquad (14)$$

to calculate the configurational probabilities, where $V_k$, $H_k$ and $S_k^{(vib)}$ are the supercell volume, the enthalpy and the vibrational entropy, respectively, for the particular configuration. In this case, the effective thermodynamic potentials for the disordered solid are:

$$H = \sum_{k=1}^{K} P_k H_k \qquad (15)$$

$$G = -k_B T \sum_{k=1}^{K} \exp(-G_k^{(vib)} / k_B T) \qquad (16)$$

and

$$S = \frac{H-G}{T} = \sum_{k=1}^{K} P_k S_k^{(vib)} - k_B \sum_{k=1}^{K} P_k \ln P_k \qquad (17)$$

where in the last expression for the entropy, the first term is the vibrational contribution and the second term is the configurational contribution. The correction defined by expression (11) can be analogously applied here to treat highly disordered systems.

## 3.3 Accessing the configurational space

Computing energies and other properties of all possible configurations of ions in a mixed solid can be a rather demanding task, even for relatively small cells. Let us consider the case of a body-centered cubic (bcc) binary alloy, with a 2×2×2 supercell, which has only 16 exchangeable sites. The number of configurations as a function of the substitution fraction $x$ increases very quickly and reaches a maximum at $x$=0.5, when there is a total of 12870 configurations (**figure 2**). This number is tractable with methods of type I, or even type II, but already becomes too expensive for type III methods. In a 3×3×3 supercell, with 54 exchange sites, the maximum is  ~$2 \times 10^{15}$ configurations, which becomes very expensive even for type I methods.

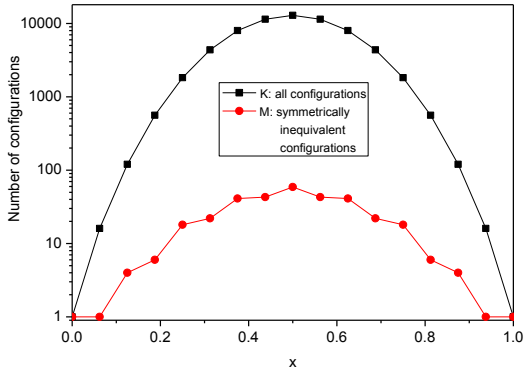

**Figure 2.** Number of configurations ($K$) in a model of a binary alloy with bcc structure using a 2x2x2 supercell, in comparison with the number of symmetrically inequivalent configurations.

It is therefore clearly necessary to find strategies to reduce the number of configurations to evaluate. We will discuss here three possible routes: *i)* taking advantage of the crystal symmetry, *ii)* random sampling, and *iii)* importance sampling using Metropolis – Monte Carlo algorithms.

### 3.4 Taking advantage of the crystal symmetry

If we are dealing with relatively small supercells, for example, when doing quantum-mechanical calculations for each configuration, it is possible to reduce the number of configurations by taking advantage of the crystal symmetry of the lattice [9]. Within this approach, two configurations are considered equivalent when they are related by a symmetry (an isometric) operation, for example, a reflection. A list of all possible isometric transformations is provided by the group of symmetry operators in the parent structure (the original structure without any substitutions). They include the symmetry operators in the space group of the crystal unit cell (scaled in an appropriate way to account for the cell multiplicity of the supercell), the supercell internal translational operators, and the combinations between them. It is then possible (at least for small systems) to start with all possible configurations through explicit enumeration and reduce to those which are symmetrically inequivalent for energy/properties evaluation. We also need to keep track of the degeneracy of each independent configuration, that is, how many times it repeats in the whole configurational space (this is similar to a number of members in the star of k-points in the Brillouin zone in the representation and group theory of crystals). This algorithm is implemented in the SOD (Site Occupancy Disorder) program [9,10].

It is necessary to slightly adapt the equations for configurational statistics to operate in the reduced space of inequivalent configurations. If $E_m$ is the energy and $\Omega_m$ is the degeneracy of the independent configuration $m$ ($m$=1,…, $M$), its contribution to energy or other properties needs to be weighted by a probability:

$$P_m = \frac{\Omega_m}{Z} \exp(-E_m / k_\mathrm{B}T) \tag{18}$$

which means that if we want to compare the stability of two independent configurations in energetic terms, we should not use their energies $E_m$ but instead the value $E_m - k_B T \ln \Omega_m$. Average values can be obtained using the equation analogous to (4):

$$A = \sum_{m=1}^{M} P_m A_m \tag{19}$$

For scalar properties, *i.e.* if $A_m$ is the same for all the $\Omega_m$ equivalent configurations that the inequivalent configuration $m$ represents. For example, if we are modeling a cubic system, we cannot obtain the average cell parameter $a$ from the cell parameters $a_m$ of the inequivalent configurations, as this result could be different from the direct average of the $b_m$ or $c_m$ values, breaking the cubic symmetry. We therefore need to find first a related magnitude that is invariant in the subspace of equivalent configurations, *e.g.* the volume $V_\mathrm{m}$ in the given example. We can then define the average cell parameter of the cubic system as:

$$a = \left( \sum_{m=1}^{M} P_m V_m \right)^{1/3}$$

(20)

Otherwise, one needs to explicitly symmetrize the property obtained using Eqn. (19) with all the symmetry operations used in obtaining the reduced set of configuration (for example, electric dipole or polarization in a ferroelectric).

Symmetry reduction is only practical when working with relatively small supercells. This is typically the case when properties other than the energy are being evaluated, using type II and type III methods. When evaluating the thermodynamic functions of the solution, the enthalpy tends to converge very quickly with supercell size, but the convergence of the entropy, which depends on configuration counting, is much slower. Therefore, for a complete thermodynamic characterization of solid solutions it is generally necessary to consider supercells much larger than those tractable by symmetry-reduction methods.

## 4. Examples of applications

### 4.1 Using symmetry-adapted ensembles to identify favourable ion distributions

We start with the simplest possible use of the multi-configurational representation of the ionic distribution in a mixed solid: finding the most stable configurations. We use the iron oxide γ-$Fe_2O_3$ (maghemite) as a first example.

Maghemite is the second most stable polymorph of iron (III) oxide. Its magnetism, chemical stability and low cost led to its wide application as magnetic pigment in electronic recording media since the late 1940's [11]. Maghemite nanoparticles are also widely used in biomedicine, because their high magnetic moment allows manipulation with external fields, while they are biocompatible and potentially non-toxic to humans [12,13]. Despite the compositional simplicity, its precise structure has been the subject of debate for decades. Like magnetite ($Fe_3O_4$), maghemite exhibits a spinel crystal structure, but while the former contains both $Fe^{2+}$ and $Fe^{3+}$ cations, in maghemite all the iron cations are in trivalent state, and the charge neutrality of the cell is guaranteed by the presence of cation vacancies. The debate about the maghemite structure has focused on the degree of ordering of these vacancies in the solid.

The unit cell of magnetite can be represented as $(Fe^{3+})_8[Fe^{2.5+}]_{16}O_{32}$, where the brackets () and [] designate tetrahedral and octahedral sites, respectively. The maghemite structure can be obtained by creating 8/3 vacancies out of the 24 Fe sites in the cubic unit cell of magnetite. These vacancies are known to be located in the octahedral sites [14] and therefore the structure of maghemite can be approximated as a cubic unit cell with composition $(Fe^{3+})_8[Fe^{3+}_{5/6}\Box_{1/6}]_{16}O_{32}$. If the cation vacancies were randomly distributed over the octahedral sites, as it was initially assumed, the space group would be Fd3m like in magnetite. However, there is a evidence about a higher degree of ordering. Braun [15], for example, noticed that maghemite exhibits the same superstructure as lithium ferrite ($LiFe_5O_8$), which is also a spinel with unit cell composition $(Fe^{3+})_8[Fe^{3+}_{3/4}Li^{1+}_{1/4}]_{16}O_{32}$, and suggested this was due to similar ordering in both compounds. In the space group $P4_332$ of lithium ferrite, there are two types of octahedral sites, one with multiplicity 12 in the unit cell, and one with multiplicity 4, which is the one occupied by Li. In maghemite, the same symmetry exists if the Fe vacancies are constrained to these Wyckoff 4b sites, instead of being distributed over *all* the 16 octahedral sites. It should be noted, however, that some level of disorder persists in this structure, as the 4b sites have fractional (1/3) iron occupancies. Finally, there is also evidence of a fully ordered structure, exhibiting a tetragonal cell with space group $P4_12_12$ and $c/a \approx 3$ (spinel cubic cell tripled along the *c* axis) [16,17].

A computational investigation of the energetics of vacancy ordering in maghemite was presented in [18]. A 1x1x3 supercell of the cubic structure was used to obtain the spectrum of energies of all the ordered configurations which contribute to the partially disordered $P4_332$ cubic structure. The energies were evaluated using long-range Coulomb contributions and classical interatomic potentials to describe short-range interactions (parameters derived by Lewis and Catlow[19]). The core-shell model of Dick and Overhauser[20] was employed to account for the polarizability of the anions. The calculations were performed with the GULP code [1,2,21]. Although not as sophisticated as quantum mechanical calculations, this methodology allows for accurate calculations of ion relaxations and configuration energies.

The total number of combinations of the 4 Fe ions on the so-called L sites of the supercell (**figure 3a**) is $12!/(4! \times 8!)=495$, but only 29 of these are inequivalent, as determined using the SOD program [9]. The calculated energies for these 29 configurations is shown in Fig. 4b. Only one of these configurations has the space group $P4_12_12$, found by Shmakov *et al.* [16] for fully ordered maghemite. This configuration is indeed the most stable one, with a significant energetic separation from the second most stable configuration (32 kJ/mol). The energy range covered by the configurational spectrum is quite wide (~850 kJ/mol), indicating that full disorder is very unlikely. The distinctive feature of the most stable configuration ($P4_12_12$) is the maximum possible homogeneity of iron cations and vacancies over the L sites. This configuration is the only one in which vacancies never occupy three consecutive layers; there are always two layers containing vacancies separated by a layer without vacancies, which instead contains $Fe^{3+}$ cations in the L sites (*e.g.* positions L1 - L4 - L7 - L10) and the $P4_12_12$ configuration is therefore the one that minimizes the electrostatic repulsion between these cations.

In order to interpret the energy differences in the configurational spectrum in terms of the degree of vacancy ordering in the solid, we can calculate the probability of occurrence of each independent configuration. **Figure 4** shows the probabilities of the most stable configuration ($P4_12_12$) and of the second most stable configuration (with space group $C222_1$) as a function of temperature. At 500 K, a typical synthesis temperature for maghemite [16], the cumulative probabilities of all the configurations excluding the most stable $P4_12_12$ is less than 0.1%. This contribution increases slowly with temperature, but at 800 K this cumulative probability, which measures the expected level of vacancy disorder, is still less than 2%. At temperatures above 700-800 K maghemite transforms irreversibly to hematite ($\alpha$-$Fe_2O_3$), and considering higher temperatures is therefore irrelevant. It thus seems clear that perfect crystals of maghemite in configurational equilibrium should have a fully ordered distribution of cation vacancies. Further analysis of the cation distribution in this oxide can be found in Ref. [18].

## 4.2 Configurational averages in the bulk and the surfaces: $Ce_{1-x}ZrO_2$ solid solutions

We discuss now some applications of the concept of configurational average, using the $Ce_{1-x}ZrO_2$ solid solution as a case study. This material is used as a support for the noble metals in the catalyst employed for the reduction of harmful emissions from car exhausts. A computational study of this solid solution was presented in ref. [22].

A supercell with 36 atoms was used there to model the bulk system, in particular the Ce-rich part of the solid solution ($0<x<0.5$ in $Ce_{1-x}Zr_xO_2$), which exhibits cubic symmetry ([23,24]). In this case, all calculations were performed using quantum-mechanical calculations, based on the density functional theory (DFT), as implemented in the VASP code [25,26]. From the calculations, it was immediately clear that the lowest-energy configurations were those where all the Zr ions are grouped together, indicating a tendency to ex-solution. The tendency to ex-solution within bulk phases can be quantified by calculating the enthalpy of mixing:

$$\Delta H_{mix} = H[Ce_{1-x}Zr_xO_2] - (1-x)H[CeO_2] - xH[\text{c-}ZrO_2]$$

$$(21)$$

where $H[CeO_2]$ and $H[\text{c-}ZrO_2]$ are the DFT energies per formula unit of ceria and cubic zirconia, respectively, and $H[Ce_{1-x}Zr_xO_2]$ is the effective energy of the solid solution, calculated as a configurational average. The resulting enthalpy of mixing is strongly positive, in agreement with recent calorimetric measurements[24] (**Figure 5**).

Assuming a regular solid solution model (*e.g.*[27, 28]), the enthalpy of mixing at low Zr content was fitted with a polynomial of the form:

$$\Delta H_{\mathrm{mix}} = Wx(1 - x)$$

$$(22)$$

as in previous experimental work ([24,29]), which gives $W=38$ kJ/mol. This result is intermediate between the value of 28 kJ/mol obtained by [29]) from fitting a regular solution model to experimental solubility data, and the value of 51 kJ/mol obtained by [24]) by fitting directly to calorimetric measurements. The positive values of the enthalpy of mixing suggest that cation ordering is not a stabilizing factor in ceria-zirconia solid solutions, at least for the compositions examined here, and confirm that the Zr ions have an energetic preference to segregate or form a separate Zr-rich phase. The origin of this tendency to is the difference between the ionic radii of the cations ($r[Ce^{4+}]=0.97$ Å and $r[Zr^{4+}]=0.84$ Å, for 8-fold coordination, according to [30]). It should be noted that real samples, where homogeneity at the atomic level can be achieved using special synthesis methods (*e.g.*[23,31]), might not experience this trend unless subjected to temperatures high enough to overcome the cation diffusion barriers.

In order to describe the thermodynamic stability of the solid solution at any finite temperature, entropies and free energies of mixing should be also calculated. It was found that, even assuming ideal configurational entropy, the resulting free energy of mixing is positive except for very small values of *x*. Furthermore, since Zr-rich phases are known to be monoclinic ([32]) at the temperatures of interest here, the mixing free energy should be calculated with respect to the more stable monoclinic zirconia phase (m-$ZrO_2$), which makes the mixed phase even less stable with respect to phase

separation. In order to estimate the solubility limit of Zr in $CeO_2$, the mixing free energy function:

$$\Delta G_{\mathrm{mix}}(x,T) = Wx(1-x)$$
$$+\Delta H_t x + RT[x\ln x + (1-x)\ln(1-x)]$$

$$(23)$$

was considered, where the enthalpy of the monoclinic-cubic zirconia phase transformation $\Delta H_t =8.8$ kJ/mol[33] was introduced. The use of the ideal entropy is justified because at very low Zr content the disorder should be nearly perfect. This analytical function allows the interpolation to *x* values smaller than those directly obtainable with the simulation supercell, and its minimum with respect to *x* at a given temperature provides an estimation of the solubility limit. **Figure 6** shows that the maximum equilibrium solubility of Zr from monoclinic zirconia into the ceria structure is ~0.4 mol% at 973 K, and increases to 2 mol% at 1373 K. Thus, although ceria-zirconia solid solutions in the whole range of compositions can be synthesized under adequate conditions (*e.g.*[31]), these results taken together with previous experimental evidence clearly show that these solid solutions are metastable with respect to phase separation into Ce-rich and Zr-rich phases. This phase separation can actually occur in a close-coupled catalytic converter, where temperatures of up to 1373 K could lead to rearrangement of the cations in the solid solution.

Simulations of the distribution of cations near the (111) surface of the solid were performed in the same study, by using the periodic slab model shown in Fig. 8. The number of configurations in the slab was reduced by only including those keeping the inversion symmetry of the cell and then selecting the symmetrically inequivalent ones. The equilibrium zirconium content of a particular cation layer parallel to the (111) surface, depends both on the overall zirconium content of the slab and on the temperature, and can be calculated by taking the configurational average:

$$c_l = \frac{\sum_m f_{ml} \Omega_m \exp(-E_m / k_B T)}{\sum_m \Omega_m \exp(-E_m / k_B T)}$$

(24)

where $f_{ml}$ is the fraction of sites occupied by Zr in the layer $l$ for configuration $m$. The results are shown in Fig. 8 for temperatures between 800 and 1600 K. The most obvious feature of the cation distribution is the low concentration of Zr at the top (111) layer. Even for the 50:50 solid solution, at the highest temperature considered (1600 K), the equilibrium Zr content of the surface is only ~10%. The dependence of the calculated concentrations on temperature is relatively weak, especially at the top layer, but it is clear that increasing temperatures lead to more homogeneity in the composition of the interior of the slab, by equalizing the Zr content in the second and third layers. Thus, according to these results, the redistribution of cations at high temperatures should occur with significant Ce-enrichment of the (111) surface of ceria-zirconia, regardless of the overall composition of the solid solution. These conclusions are discussed in detail, in comparison with the experimental evidence, in ref. [22]

### 4.3 Stability of titanium oxynitrides

TiN and TiO2 (the most stable nitride and oxide phases of titanium) have an impressive number of interesting properties and potential applications in key technological fields. However, the properties are very different from one to other, and a complete change in the electronic and geometric structures takes place when TiN is oxidized to $TiO_2$ and when $TiO_2$ is nitrided to TiN. A number of intermediate phases of general composition $TiO_xN_y$, called "oxynitrides", appear in these complex processes. Obviously, the properties of the oxynitrides will be similar to those of the respective pure nitride and oxide when their compositions are close to those of the pure systems, and they will change progressively from those of the nitride to those of the oxide and vice versa when the compositions move to intermediate values. In principle, one could control and modulate the properties of the system by controlling the composition of the oxynitrides. In that way, potentially interesting combined properties could be obtained. But there are a number of questions to solve: are those oxynitrides stable phases which we are able to synthesize? What are their structures? Are their properties a result of the combination of those of the pure solids? Are we really able to control these properties as a function of the composition $TiO_xN_y$?

In order to answer these questions we made use of the SOD code. We performed[34] DFT calculations using a Generalized Gradient Approximation (GGA) implemented in the VASP code.[25] A plane-wave cutoff energy of 500 eV was used. We chose a supercell model of (1x1x3) for TiN (24 atoms) and one of (1x1x1) for α-TiO (20 atoms). These models allow us to change the composition progressively while still having a computationally affordable size, since we optimize the geometries with high accuracy (cutoff of 500 eV, saturation of k-points and demanding convergence criterions) for all the possible different configurations for each composition (240 configurations). The calculations were carried out using a (7x7x3) mesh for TiN and oxynitrides with NaCl-type structure, and a (6x4x6) mesh for TiO and oxynitrides with α-TiO-type structure.

We performed an exhaustive study of all the possible configurations of the systems, i.e. we studied all the different arrangements of the N and O atoms in the unit cells. For example, if we want to model a $TiN_{1-x}O_x$ system with NaCl-type structure, in which the N/O ratio is 0.5, we use the supercell (1x1x3) for TiN (which has 24 atoms), and we substitute 6 out of the 12 N atoms by O atoms. The number of different possibilities in which we can carry out the 6 substitutions is 940. In principle, if we wanted to make sure that we have found the most stable configuration of the system $TiN_{0.5}O_{0.5}$ we should perform the 940 geometry optimisations, which would be an impossible task, given the current computer time limitations. In order to perform the exhaustive study of all the configurations, while still using an acceptable amount of computer resources, we employed the SOD code, which makes use of the symmetry of the system to reduce drastically the number of configurations.

In the previous example, most of the 940 configurations are found to be equivalent. Two configurations are equivalent when they are related by an isometric operation (such as translations, rotations or reflections within the supercell, which are consistent with the symmetry operations of the crystal). Using the SOD code to remove the equivalent configurations we find that the number of non-equivalent configurations of the cited example is only 34, which is tractable with our computer resources. Employing the SOD code we also performed a statistical analysis of all the possible configurations of the $TiN_{0.5}O_{0.5}$ system, with which we obtained the energy of the system as a weighted average of the 940 configurations.

In order to study the $TiN_{1-x}O_x$ systems with NaCl-type structures, for $x$=0, 0.16, 0.33, 0.5, 0.66, 0.83 and 1, we substituted respectively 0, 2, 4, 6, 8, 10 and 12 of the 12 N atoms in the NaCl-type TiN supercell by O atoms. Using the SOD code we calculated the number of non-equivalent configurations, which is 1, 5, 21, 34, 21, 5 and 1 respectively. The total number of configurations we studied with the NaCl-type structure is therefore 88.

In the case of the $TiN1-xOx$ systems with alpha-TiO structures, the supercell had 10 Ti atoms and 10 N or O atoms. The concentrations studied are x=0, 0.2, 0.4, 0.6, 0.8 and 1, which are achieved by substituting 0, 2, 4, 6, 8 and 10 of the 10 O atoms in the alpha-TiO supercell by N atoms. The number of non-equivalent configurations in this case is 1, 15, 60, 60, 15 and 1 respectively, giving a total number of 152. Note that, even with a smaller number of atoms in the supercell (20 as opposed to 24), the number of non-equivalent configurations in the case of the α-TiO structure is almost twice as large as that in the case of the NaCl-type structure. The reason of that is the high symmetry of the latter structure, which allows a great reduction of the number of configurations. The structural evolution of $TiN_{1-x}O_x$ compounds has been studied through the evolution of the formation energy with both the composition (x) and the structure (NaCl or α). The formation energy was calculated as follows:

$$E_f\left(TiN_{1-x}O_x\right) = E\left(TiN_{1-x}O_x\right) - nE(Ti\,\text{bulk})$$
$$- \frac{n(1-x)}{2}E(N_2) - \frac{nx}{2}E(O_2) \qquad (25)$$

Where $n$ is the number of Ti atoms, E(Ti bulk) is the energy of the bulk of metallic Ti per Ti atom, and E($N_2$) and E($O_2$) are the energies of the isolated $N_2$ and $O_2$ molecules respectively.

Obviously, when $x$ is close to 0 the system will have tendency to arrange itself as NaCl-type structure, since that is the most stable structure for TiN. Analogously, when x is close to 1 the system will try to arrange itself as alpha structure since this is the most stable structure for TiO. What we have to calculate is the limit composition at which the change of crystal structure takes place, and whether this limit composition depends on the temperature. **Figure 8** shows the evolution of the formation energy with the composition for both structures at 10 K. As it was predicted to happen, the NaCl-type structure is the preferred one for compositions close to $x$=0 (TiN), while the alpha structure is the most stable for compositions close to $x$=1 (TiO). The crossing point is found to be at the limit composition of $x$=0.55-0.60, approaching 0.60 as the temperature increases (curves were calculated at 10 K, 300 K and 600 K). The system will tend to acquire the NaCl-type structure for compositions x<0.6 while it will try to be ordered as alpha structure for compositions x>0.6. In the later case one should expect to find a number of vacancies in both Ti and N/O sublattices, since this is one of the main characteristics of the alpha structure. This is important from a technical point of view, since the presence of vacancies may change drastically the surface stability of the solid and generate highly reactive surfaces, which would be a serious drawback for microelectronic devices or technologies based on thin-films, but it could become interesting from a chemical point of view.

## 5. Conclusions

Site disorder is an important phenomenon that affects the structure and properties of materials. We have reviewed here some strategies for modeling and simulations to capture the physics and chemistry of disorder in influencing various properties of materials. These typically involve access to configurational information at different levels, like electronic properties, atomic displacements, and have varied computational cost. While cluster expansions have been used extensively in determination of phase diagrams of alloys and similar problems, we have emphasized here the methods that attempt essentially an exact statistical thermodynamic analysis using the SOD technique with a relatively smaller system, but having access to as much information and properties as possible. Such an approach is becoming quite practical in understanding and design of disordered materials, thanks to advances in computers and algorithms.

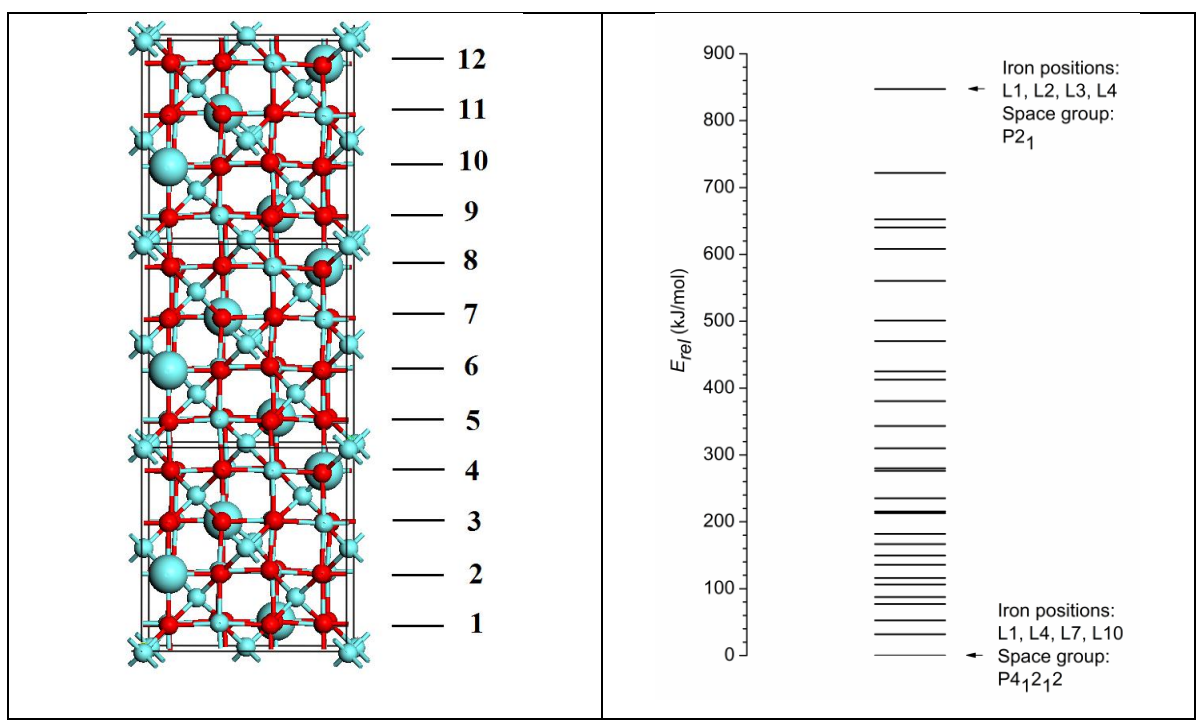

**Figure 3.** a) The exchangeable sites in the maghemite tetragonal cell: 4 Fe ions and 8 vacancies are distributed over these "L" sites. b) The calculated configurational spectrum.

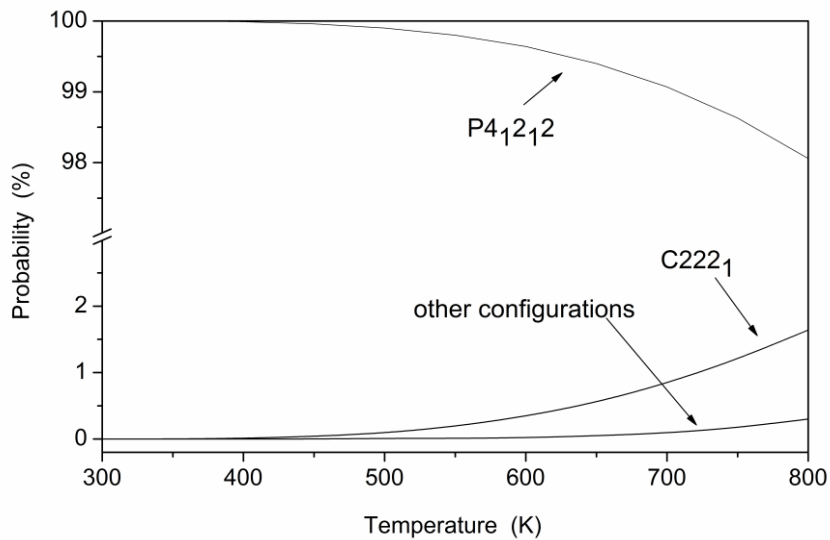

**Figure 4.** Probabilities of the two most stable configurations in the maghemite supercell as a function of temperature.

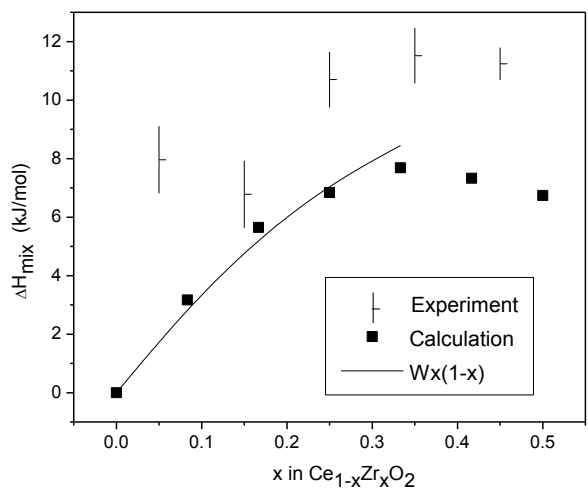

**Figure 5.** Calculated enthalpies of mixing for $Ce_{1-x}Zr_xO_2$ in comparison with experimental results [24]. The curved line represents the fitting of a regular-solution quadratic polynomial to the calculated values for low Zr concentrations.

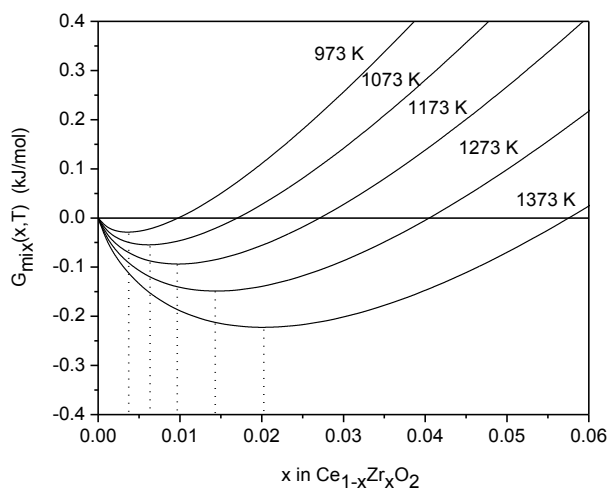

**Figure 6.** Free energies of mixing for low Zr concentrations. The vertical dotted lines mark the solubility limit of Zr in CeO$_2$ at the particular temperature.

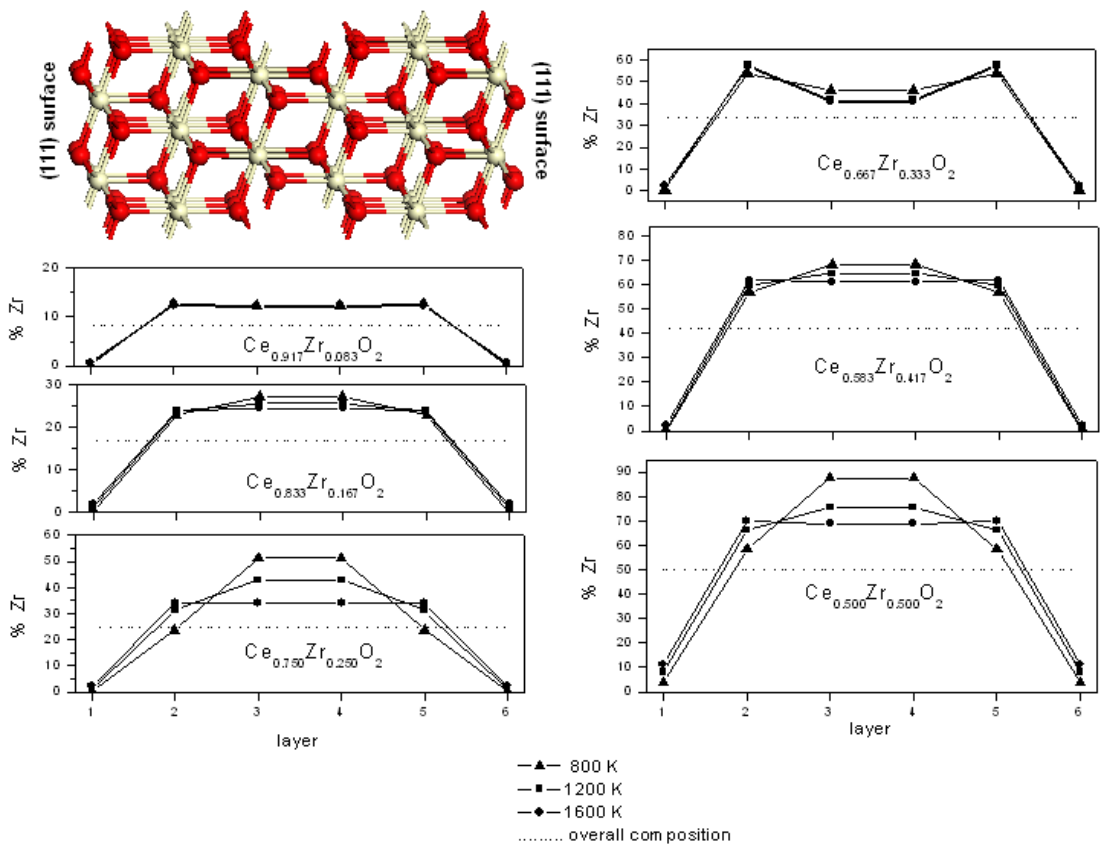

**Figure 7.** Calculated equilibrium concentrations of Zr as a function of the distance to the (111) surface in the Ce$_{1-x}$Zr$_x$O$_2$ solid solution. Because of the slab construction, layers 1, 2 and 3 are equivalent to layers 6, 5 and 4, respectively.

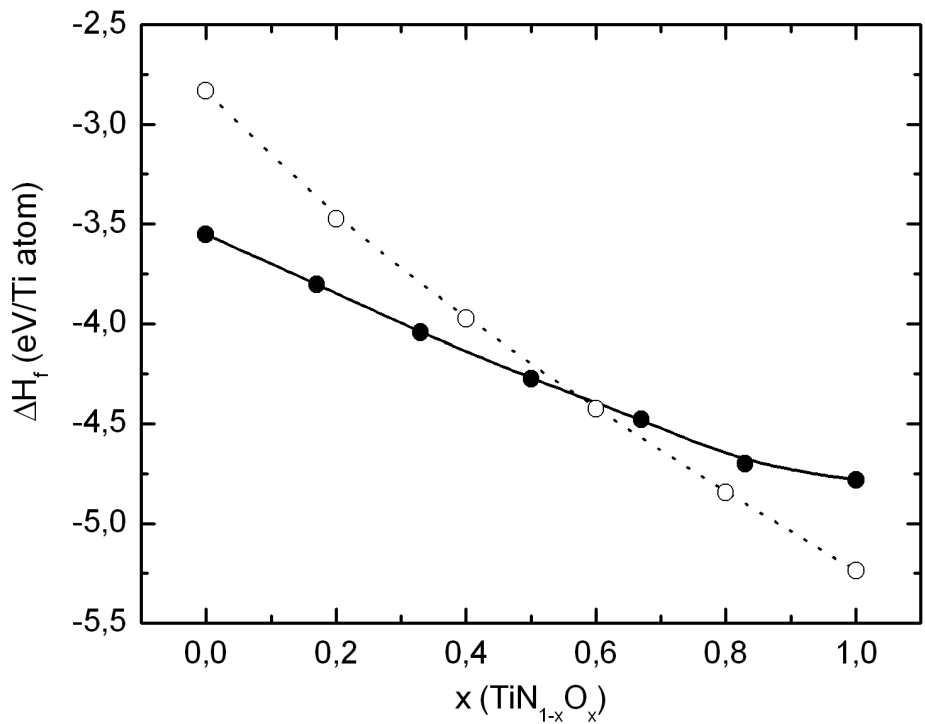

**Figure 8.** Evolution of the formation energy per Ti atom (eV) with the composition for the both structures NaCl-type (black circles) and α-type (white circles).

## Acknowledgments

We are very grateful to all our co-authors in the case studies shown here, and in particular to Prof. Umesh Waghmare, Prof. Richard Catlow, and Prof. Nora de Leeuw for their contributions to the development of the ideas presented in this summary.

## References and Notes

1 Gale, J. D. GULP: A computer program for the symmetry-adapted simulation of solids. *J. Chem. Soc.-Faraday Trans.* **93**, 629-637 (1997).
2 Gale, J. D. & Rohl, A. L. The General Utility Lattice Program (GULP). *Mol. Simul.* **29**, 291-341 (2003).
3 Zunger, A., Wei, S. H., Ferreira, L. G. & Bernard, J. E. Special quasirandom structures. *Physical Review Letters* **65**, 353 (1990).
4 Catlow, C. R. A. *Computer Modelling in Inorganic Crystallography*. (Academic Press Limited 1997).
5 van de Walle, A. & Ceder, G. The effect of lattice vibrations on substitutional alloy thermodynamics. *Reviews of Modern Physics* **74**, 11 (2002).
6 Becker, U., Fernandez-Gonzalez, A., Prieto, M., Harrison, R. & Putnis, A. Direct calculation of thermodynamic properties of the barite/celestite solid solution from molecular principles. *Physics and Chemistry of Minerals* **27**, 291-300 (2000).

7       Dove, M. T. *Introduction to Lattice Dynamics*.  (Cambridge University Press, 1993).

8       Benny, S., Grau-Crespo, R. & De Leeuw, N. H. A theoretical investigation of $\alpha$-$Fe_2O_3$– $Cr_2O_3$ solid solutions. *Physical Chemistry Chemical Physics* **11**, 808 - 815 (2009).

9       Grau-Crespo, R., Hamad, S., Catlow, C. R. A. & de Leeuw, N. H. Symmetry-adapted configurational modelling of fractional site occupancy in solids. *Journal of Physics-Condensed Matter* **19**, 256201 (2007).

10      Grau-Crespo,     R.     &     Hamad,     S.     *SOD     (Site-Occupancy     Disorder)*, <https://sites.google.com/site/rgrauc/sod-program> (2007).

11      Dronskowski, R. The little maghemite story: A classic functional material. *Advanced Functional Materials* **11**, 27-29 (2001).

12      Pankhurst, Q. A., Connolly, J., Jones, S. K. & Dobson, J. Applications of magnetic nanoparticles in biomedicine. *J. Phys. D-Appl. Phys.* **36**, R167-R181 (2003).

13      Levy, M. *et al.* Magnetically induced hyperthermia: size-dependent heating power of gamma-Fe2O3 nanoparticles. *Journal of Physics-Condensed Matter* **20** (2008).

14      Waychunas, G. A. Crystal chemistry of oxides and oxyhydroxides. *Reviews in Mineralogy and Geochemistry* **25**, 11-68 (1991).

15      Braun, P. B. A superstructure in spinels. *Nature* **170**, 1123 (1952).

16      Shmakov, A. N., Kryukova, G. N., Tsybulya, S. V., Chuvilin, A. L. & Solovyeva, L. P. Vacancy Ordering in Gamma-Fe2o3 - Synchrotron X-Ray-Powder Diffraction and High-Resolution Electron-Microscopy Studies. *J Appl Crystallogr* **28**, 141-145 (1995).

17      Jorgensen, J. E., Mosegaard, L., Thomsen, L. E., Jensen, T. R. & Hanson, J. C. Formation of gamma-Fe2O3 nanoparticles and vacancy ordering: An in situ X-ray powder diffraction study. *Journal of Solid State Chemistry* **180**, 180-185 (2007).

18      Grau-Crespo, R., Al-Baitai, A. Y., Saadoune, I. & De Leeuw, N. H. Vacancy ordering and electronic structure of $\gamma$-Fe2O3 (maghemite): a theoretical investigation. *Journal of Physics: Condensed Matter* **22**, 255401 (2010).

19      Lewis, G. V. & Catlow, C. R. A. Potential Models for Ionic Oxides. *Journal of Physics C-Solid State Physics* **18**, 1149-1161 (1985).

20      Dick, B. G. & Overhauser, A. W. Theory of the dielectric constant of alkali halide crystals. *Physical Reviews* **112**, 90-103 (1958).

21      Gale, J. D. GULP: Capabilities and prospects. *Z. Kristallogr.* **220**, 552-554 (2005).

22      Grau-Crespo, R., De Leeuw, N. H., Hamad, S. & Waghmare, U. V. Phase separation and surface segregation in ceria-zirconia solid solutions. *Proceedings of the Royal Society A-Mathematical     Physical     and     Engineering     Sciences*,     DOI:10.1098/rspa.2010.0512 doi:10.1098/rspa.2010.0512 (2011).

23      Cabanas, A., Darr, J. A., Lester, E. & Poliakoff, M. Continuous hydrothermal synthesis of inorganic materials in a near-critical water flow reactor; the one-step synthesis of nano-particulate $Ce_{1-x}Zr_xO_2$ (x=0-1) solid solutions. *J. Mater. Chem.* **11**, 561-568 (2001).

24      Lee, T. A., Stanek, C. R., McClellan, K. J., Mitchell, J. N. & Navrotsky, A. Enthalpy of formation of the cubic fluorite phase in the ceria–zirconia system. *Journal of Materials Research* **23**, 1105-1112 (2008).

25      Kresse, G. & Furthmuller, J. Efficiency of ab-initio total energy calculations for metals and semiconductors using a plane-wave basis set. *Comput. Mater. Sci.* **6**, 15-50 (1996).

26      Kresse, G. & Furthmuller, J. Efficient iterative schemes for ab initio total-energy calculations using a plane-wave basis set. *Phys. Rev. B* **54**, 11169-11186 (1996).

27      Prieto, M. Thermodynamics of Solid Solution-Aqueous Solution Systems. *Thermodynamics and Kinetics of Water-Rock Interaction* **70**, 47-85 (2009).

28      Ruiz-Hernandez, S. E., Grau-Crespo, R., Ruiz-Salvador, A. R. & De Leeuw, N. H. Thermochemistry of strontium incorporation in aragonite from atomistic simulations. *Geochimica Et Cosmochimica Acta* **74**, 1320-1328 (2010).

29    Du, Y., Yashima, M., Koura, T., Kakihana, M. & Yoshimura, M. Thermodynamic evaluation of the $ZrO_2$–$CeO_2$ system. *Scripta Metall. Mater.* **31**, 327 (1994).

30    Shannon, R. D. Revised effective ionic radii and systematic studies of interatomic distances in halides and chalcogenides. *Acta Crystallographica* **A32**, 751-767. (1976).

31    Cabanas, A., Darr, J. A., Lester, E. & Poliakoff, M. A continuous and clean one-step synthesis of nano-particulate $Ce_{1-x}Zr_xO_2$ solid solutions in near-critical water. *Chemical Communications*, 901-902 (2000).

32    Garvie, R. C. in *High temperature oxides. Part II* (ed M.A. Alper) 117 (Academic Press, 1970).

33    Navrotsky, A., Benoist, L. & Lefebvre, H. Direct calorimetric measurement of enthalpies of phase transitions at 2000 degrees–2400 degrees C in yttria and zirconia. *Journal of the American Ceramic Society* **88**, 2942 (2005).

34    Graciani, J., Hamad, S. & Sanz, J. F. Changing the physical and chemical properties of titanium oxynitrides TiN(1-x)Ox by changing the composition. *Phys. Rev. B* **80**, 184112 (2009).